\documentclass[prb,twocolumn,amsmath,amssymb]{revtex4}

\usepackage{graphicx}
\usepackage{dcolumn}
\usepackage{bm}

\begin{document}

\newcommand{\up}{\uparrow}
\newcommand{\down}{\downarrow}
\newcommand{\phdagger}{\phantom{\dagger}}          
\newcommand{\aop}[1]{a^{\phdagger}_{#1}}           
\newcommand{\adop}[1]{a^{\dagger}_{#1}}            
\newcommand{\bop}[2]{b^{\phdagger}_{#1}(#2)}           
\newcommand{\bdop}[2]{b^{\dagger}_{#1}(#2)}            
\newcommand{\cop}[1]{c^{\phdagger}_{#1}}           
\newcommand{\cdop}[1]{c^{\dagger}_{#1}} 
\newcommand{\dop}[2]{d^{\phdagger}_{#1}(#2)}           
\newcommand{\ddop}[2]{d^{\dagger}_{#1}(#2)}            
\newcommand{\sS}{{\mathcal{S}}}
\newcommand{\sH}{{\mathcal{H}}}
\newcommand{\mgt}{>>}
\newcommand{\mlt}{<<}
\newcommand{\half}{\frac{1}{2}}
\newcommand{\rp}{r^{\prime}}
\def\sO{{\mathcal{O}}}
\def\cd{c^{\dagger}}
\newcommand{\bra}[1]{| #1 \rangle}


\title{Band structure of Charge Ordered Doped Antiferromagnets.}
\author{Mats Granath}
\email{mgranath@fy.chalmers.se}
\affiliation{%
Chalmers Technical University\\
G\"oteborg 41296 \\
Sweden
}

\date{\today}

\begin{abstract}
We study the distribution of electronic spectral weight in a doped antiferromagnet with various 
types of charge order and compare to angle resolved photoemission experiments on
lightly doped La$_{2-x}$Sr$_x$CuO$_4$ (LSCO) and electron doped Nd$_{2-x}$Ce$_x$CuO$_{4\pm\delta}$.  
Calculations on in-phase stripe and bubble phases for the electron doped system are both in good agreement 
with experiment including in particular the existence of in-gap spectral weight. In addition we find that for in-phase 
stripes, in contrast to anti-phase stripes, the chemical potential is likely to move with doping.
For the hole doped system we find that ``staircase'' stripes which are globally diagonal but locally vertical or horizontal   
can reproduce the photoemission data whereas pure diagonal stripes cannot. We also calculate the magnetic structure factors of such 
staircase stripes and find that as the stripe separation is decreased with increased doping these evolve from diagonal to 
vertical separated by a coexistence region.   
The results suggest that the transition from horizontal to diagonal stripes seen in neutron scattering 
on underdoped LSCO may be a crossover between a regime where the typical length of straight stripe segments  
is longer than the inter-stripe spacing to one where it is shorter and that locally the stripes are always aligned with the 
Cu-O bonds. 

\end{abstract}

\maketitle
\section{Introduction\label{intro}}

In several families of superconducting cuprates there is evidence for \textit{stripes} which are 
regularly spaced quasi-one dimensional structures where the doped charge assembles.\cite{Erica_review}  
At the same time there is a well developed theory of high-temperature superconductivity in a system of weakly coupled Hubbard 
ladders.\cite{spin_gap_proximity,Fradkin,Arrigoni} It is thus quite natural to speculate that such a theory is in fact 
realized in the cuprates. 
One obvious objection to the stripe scenario of superconductivity is the lack of convincing evidence for the existence of 
stripes in several materials given that the 
strongest evidence is found in the relatively low-T$_c$ La$_{2-x}$Sr$_x$CuO$_4$ (LSCO) family.\cite{LSCO_stripes}
From a theory point of view the lack of direct evidence is not immediately discouraging because stripe order is bad 
for superconductivity whereas more elusive dynamic charge stripe correlations are good.\cite{Kivelson,Tranquada}
However, even in a system such as LSCO where stripes are well established there are still issues about their fundamental 
implications on the electron dynamics. In particular, it seems the distribution of low-energy spectral 
weight in k-space as probed by angle-resolved photoemission spectroscopy (ARPES) does not show any clear evidence of the 
suggested quasi-one-dimensional nature of the electronic states, looking instead like the Fermi surface of a fully 
two-dimensional system.\cite{note_on_1D}  On the other hand
ARPES also provides clear evidence of exotic physics, perhaps electron fractionalization\cite{fractionalization}, 
with spectral functions that are typically very broad in energy and not 
consistent with a Fermi liquid based quasi-particle description.\cite{NFL_ARPES}

A simple model for studying the distribution of single electron spectral weight in a charge ordered doped antiferromagnet 
was introduced by Salkola et al.\cite{Salkola} in which all the complicated physics which is responsible for the 
stripe formation and integrity is replaced by a hand picked potential which emulates the local environment of 
electrons in a stripe ordered antiferromagnet. The potential is simply a staggered field representing the local 
antiferromagnetism together with anti-phase domain walls of suppressed field strength 
representing stripes. The spacings of the stripes are chosen in such a way that the model by construction will 
reproduce the diffraction response of a stripe ordered system. The stripe placements are static but may be 
chosen in an irregular fashion simulating quenched disorder or dynamic stripes which are fluctuating slowly 
compared to time scales of the local electron dynamics.
Using this model for a disordered 
array of quarter filled ``bond-aligned''\cite{bond_aligned} stripes it was found that the low energy spectral weight 
forms a two-dimensional Fermi surface with, in particular, spectral weight in the ``nodal regions'', near $(\pi/2,\pi/2)$, 
which naively corresponds to propagation diagonally with respect to the stripe direction. 
A more detailed study of this model was carried out in Ref. \onlinecite{Granath} where it was realized that 
even a single stripe with one-dimensional states localized transverse to the stripe captures the 
qualitative features of the low-energy 
spectral weight as seen in ARPES in the 
under and optimally doped LSCO with the full spectral weight of the stripe lying roughly on the diamond Fermi 
surface of a half-filled nearest neighbor tight-binding model. Here also issues of interactions 
along a stripe was addressed and it was found that the distinct non-Fermi liquid properties of an interacting 
one-dimensional electron gas may be displayed also in directions not aligned with the stripes. 

In the present paper we extend the work of Ref. \onlinecite{Salkola} and \onlinecite{Granath} to look at a broader range of 
ordered structures. This is motivated  by a 
search for a more comprehensive understanding of charge order in a doped antiferromagnet 
but also by direct or indirect observations of structures which are not consistent with bond-aligned anti-phase stripes. 
In the very lightly doped, non-superconducting, phase of LSCO  
there are diffraction patterns consistent 
with diagonal stripes\cite{Wakimoto99,Fujita} instead of the bond-aligned stripes seen at higher doping. 
The transition between bond-aligned and 
diagonal stripes coincides with the superconductor to insulator transition at doping $x\approx 0.055$, suggesting a close link between the 
two properties. Within the non-interacting model considered here we cannot directly address the 
connection between superconductivity and stripe orientation but by comparing the predicted spectral weight 
of diagonal stripes with that observed in ARPES\cite{diagonal_LSCO_ARPES} we gain some insight into the nature of 
such diagonal stripes.
Recently, there has also been indirect evidence both from NMR\cite{NMR_e_doped} and thermal conductivity 
measurements \cite{Ando} 
of an inhomogeneous charge structure in the electron-doped cuprates (Pr,La)$_{2-x}$Ce$_{x}$CuO$_4$ (PLCCO). 
However, in electron doped materials there has been no 
evidence for incommensurate magnetism\cite{Yamada} effectively ruling out the possibility of anti-phase stripes. For this 
reason we have studied in-phase structures and compared to ARPES results  
on the Nd$_{2-x}$Ce$_x$CuO$_{4\pm\delta}$ (NCCO) family.\cite{e_doped_ARPES}

Our results can be summarized as follows. As exemplified in Fig. \ref{figfull_spectral} we find that the 
spectral weight of localized states is centered on the 
Brillouin zone (BZ) diagonals, given by $\cos(k_x)+\cos(k_y)=0$, independently of the magnitude of second and 
third nearest neighbor hopping $t'$ and $t''$ as well as the shape and form of the charge ``impurity''.  
This shows why in general one may not expect any dramatic signature of stripes or other inhomogeneous 
electronic structure on the k-space distribution of low energy spectral weight, as  
this is where the Fermi surface is located within a nearly half-filled tight-binding model which is 
dominated by nearest neighbor hopping. However, as shown in Fig. \ref{fig_stair4}, \ref{fig_inphase2}, and \ref{fig_blob_dispersion},
a defining feature of the stripe states is that they have an energy which is within the Mott gap of the undoped system.
Such in-gap states appear to be a common feature of the evolution of the band structures which doping as measured in ARPES and which
we believe is a strong indication of an inhomogeneous charge distribution.\cite{topological_doping} 
 
We compare the dispersion and distribution of spectral weight of diagonal and bond-aligned stripes and find that pure diagonal stripes
cannot reproduce the characteristic ``Fermi arc'' centered around the nodal region 
which is seen in ARPES on lightly doped LSCO.
Instead we find that the spectral weight of a hole doped diagonal stripe is concentrated to the ``anti-nodal'' 
BZ regions around $(\pi,0)$ with very little weight in the nodal region. 
In addition, the band width of states on a diagonal stripe is expected to be roughly
$2|t'|\alt 0.2$eV which is inconsistent with the ARPES data where the 
band width of the in-gap states is of the order of 1eV (assuming half is seen). However, it turns out that the spectral distribution
and band width of bond-aligned stripes is qualitatively consistent with the ARPES data. 
For this reason we suggest that the diagonal stripe phase consists of stripes which are globally diagonal but locally 
bond-aligned, a caricature of which are the ``staircase stripes'' shown in Fig. \ref{figstairs}. In Fig. \ref{fig_stair3} 
is shown the low energy spectral weight which is concentrated around the nodal region in a hole doped array of staircase stripes. 

An interesting aspect of these staircase stripes is the magnetic structure factors which depend on the ratio between the length of the 
bond-aligned segments, the ``step'' length $l$, and the distance between neighboring stripes $d$. As shown in Fig. \ref{fig_stairsq} 
we find that the corresponding
Bragg peaks can be classified in three main regimes. For $l\approx d$ there are two peaks corresponding to diagonal stripes along the
$\hat{x}+\hat{y}$ direction at $(\pi\pm\delta_{\text{diag}},\pi\mp\delta_{\text{diag}})$, 
for $l\approx 2d$ the diagonal peaks coexist with four peaks at $(\pi,\pi\pm\delta_{\text{col}})$ and $(\pi\pm\delta_{\text{col}},\pi)$ 
corresponding to the response expected from bond-aligned stripes along the $\hat{x}$ and $\hat{y}$ directions, while for 
$l\gg d$ there are only the four bond-aligned peaks but they are now shifted away slightly from the square lattice axes. The qualitative 
features are very similar to what is found from neutron scattering in LSCO as a function of 
doping\cite{Fujita}, although the relative incommensurability $\delta_{\text{diag}}/\delta_{\text{col}}$ is not quite 
accurately reproduced in the coexistence regime.
Nevertheless, this suggests a scenario in which the stripes in the orthorhombic phase of LSCO are always locally bond-aligned but 
in some sense globally diagonal with a crossover as the stripe spacing is decreased with increased 
doping and not a first order transition as suggested by the neutron scattering data. 
Corroborating such a crossover scenario is the fact that the ARPES spectra evolve smoothly through
the diagonal to bond-aligned stripe transition and that the Fermi velocity in the nodal direction is roughly independent of 
doping, indicating that if the low-energy spectral weight is stripe related the local character of the stripes does not change 
dramatically with doping.\cite{diagonal_LSCO_ARPES} 
In addition, the hole mobility at moderate temperatures changes by only a factor of three
from very light ($x=0.01$) to optimal ($x=0.17$) doping which is very naturally understood 
within a stripe model in which the local character of the stripes is roughly independent of 
doping.\cite{hole_mobility}

In Fig. \ref{fig_inphase2} we show the the band structure of a system with disordered in-phase stripes and in 
Fig. \ref{fig_inphase1} the corresponding
electron doped low energy spectral weight which is in qualitative agreement with the ARPES data. The most interesting part of these
results is the evolution of spectral weight in the nodal region, where at light doping (4\%) there is in-gap spectral weight which at 
higher doping (10\%) broadens as a consequence of the shorter inter-stripe distance and reaches the Fermi surface. 
The evolution of the Fermi surface with doping can be reproduced in mean-field theory of Hubbard model with longer range hopping
by allowing for a doping dependent interaction $U$.\cite{Kusko}
The difference within a stripe model is that the low-energy states are dynamically one-dimensional, 
being localized transverse to the stripes (Fig. \ref{fig_local_state}) and the existence of in-gap spectral weight. 
In addition we find that for in-phase stripes, in contrast to anti-phase stripes, the chemical potential is likely to 
move into the upper Hubbard band with electron doping because the in-phase stripe states lie close to the upper and lower Hubbard 
bands (Fig. \ref{fig_inphase3}).

Finally, we present similar results from a calculation on a ``bubble'' phase where the doped charge is confined to small 
``0-dimensional'' droplets instead of the 1-dimensional stripes. Bubbles would arise naturally instead of stripes 
in a $t-J$ model with long-range Coulomb repulsion in the limit $t\ll J$ because of the lower magnetic energy.
Thus in the electron doped materials, which appear to have ``stronger'' 
antiferromagnetism than the hole doped materials, one may speculate that bubbles are favored over stripes. As far as the distribution
of spectral weight is concerned (Fig. \ref{fig_blobFS} and \ref{fig_blob_dispersion}) there is little qualitative difference 
between bubbles and stripes. For the bubbles we find that the nodal spectral weight broadens as the size of the bubbles 
increase with doping in analogy with the increasing density of stripes.

\section{The Model}
We will consider a tight-binding model on a square lattice with first, second and third nearest neighbor hopping together with a 
static potential which represents stripes or other charge structures.  
The Hamiltonian reads
\begin{eqnarray}
H=-&&t\sum_{\langle rr'\rangle\sigma}(c^{\dagger}_{r,\sigma}c_{r'\sigma}
+{\mbox{H.C.}})\nonumber\\
-&&t'\sum_{\langle rr'\rangle'\sigma}(c^{\dagger}_{r,\sigma}c_{r'\sigma}
+{\mbox{H.C.}})\nonumber\\
-&&t''\sum_{\langle rr'\rangle''\sigma}(c^{\dagger}_{r,\sigma}c_{r'\sigma}
+{\mbox{H.C.}})\nonumber\\
+&&m\sum_{x,y,\sigma}\sigma(-1)^{x+y}V(x,y)c^{\dagger}_{x,y,\sigma}
c_{x,y,\sigma} \; ,\label{H}
\end{eqnarray}
where $c_{r,\sigma}$ is the electron destruction operator at site 
$r=(x,y)$ and with spin $\sigma=\pm$. The hopping is given in a standard fashion where 
$\langle rr'\rangle$ indicates nearest neighbors, $\langle rr'\rangle'$ next-nearest neighbors, and 
$\langle rr'\rangle''$ next-next nearest neighbors. In what follows we will use energy units such that 
$t=1$ and we take $t'/t<0$ and $t''/t>0$. Unless stated otherwise we will be using $m=t=1$.
The physical intuition for the field $m$ is that it is the energy cost of moving a hole from a stripe into the antiferromagnetic 
background and it is thus expected to be of the order of the spin exchange $J$. 
We also use units such that the lattice constant is equal to one and $\hbar=1$.
  
The potential $m(-1)^{x+y}V(x,y)$ describes the collective field which defines the stripe order. We will use the simplest 
possible form and take $V(x,y)=$1,0 or -1, where $V(x,y)=$1 or -1 represent antiferromagnetic regions related by a $\pi$ phase 
shift and $V(x,y)=0$ are the locations of the stripes.
The case $V(x,y)=1$ for all $x$ and $y$ corresponds to the standard mean-field result of the Hubbard model at half-filling\cite{Schrieffer} 
giving an upper and lower Hubbard band separated by the Mott-Hubbard gap.
Introducing regions where $V(x,y)=0$ will in general give rise to localized ``impurity'' states within the gap.

\begin{figure}[h]
\includegraphics[scale=.4]{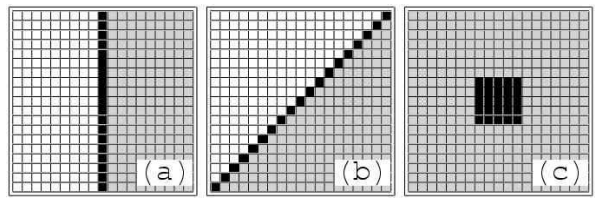}
\caption{\label{figvxy} Graphic representation of the potential $V(x,y)$ for various charge structures. Here black 
corresponds to $V=0$, gray to $V=1$, white to $V=-1$ and each square represents a site $(x,y)$ on the lattice. 
(a) is a bond-aligned anti-phase stripe, (b) is an anti-phase diagonal stripe and (c) a bubble.}
\end{figure}

The object which was studied in detail in Ref \onlinecite{Granath} is the site-centered anti-phase stripe given by
\begin{equation}
V_{\text{stripe}}(x,y)=\left\{
\begin{matrix}
1, & x>0\\
0, & x=0\\
-1,& x<0\\
\end{matrix},\right.
\end{equation}
and displayed graphically in Fig. \ref{figvxy}a. Here we also show two other charge structures, diagonal stripes and bubbles, 
which we will consider in more detail subsequently. In a real system we want to consider ordered or disordered arrays 
structures such that at some finite doping the corresponding states are partly occupied.

\subsection{Spectral weight of localized states\label{sec:analytic}}

Here we will show that a state localized on an impurity where the staggered field is zero has its 
spectral weight centered within a region $m$ of the $\cos(k_x)+\cos(k_y)=0$ diamond, 
regardless of the values of the longer range hopping $t'$ and $t''$ and the geometry of the impurity. 
This result is really a trivial consequence of the fact that the nearest neighbor hopping $t$ connects the two sublattices of 
the staggered field while $t'$ and $t''$ does not, or equivalently the symmetry or not of the dispersion with respect to a shift 
of the momenta by scattering vector $(\pi,\pi)$ of the staggered field.   

We define a potential for an arbitrary impurity 
\begin{eqnarray}
V(x,y)=& m(-1)^{x+y}\,,\quad & \{x,y\}\not\in\text{impurity}\nonumber\\
V(x,y)=& 0\,,\quad & \{x,y\}\in\text{impurity}\,,
\end{eqnarray}
and write the tight-binding dispersions 
\begin{eqnarray}
\varepsilon_{\vec{k}}&=&\varepsilon^0_{\vec{k}}+\varepsilon^1_{\vec{k}}\nonumber\\
\varepsilon^0_{\vec{k}}&=&-2t(\cos(k_x)+\cos(k_y))\\
\varepsilon^1_{\vec{k}}&=&-4t'\cos(k_x)\cos(k_y)-2t''(\cos(2k_x)+\cos(2k_y))\,,\nonumber 
\end{eqnarray}
With this we solve for the eigenstates in the bulk
\begin{equation}
\psi_{\vec{k}}=a_{\vec{k}}e^{i\vec{k}\cdot\vec{x}}+b_{\vec{k}}e^{i(\vec{k}+\vec{\pi})\cdot\vec{x}}\,,
\end{equation}
with energy
\begin{equation}
E_{\vec{k}}=\varepsilon^1_{\vec{k}}\pm\sqrt{(\varepsilon^0_{\vec{k}})^2+m^2}\,,
\end{equation}
and ratio of coefficients
\begin{equation}
\label{bovera}
b_{\vec{k}}/a_{\vec{k}}=(-\varepsilon^0_{\vec{k}}\pm\sqrt{(\varepsilon^0_{\vec{k}})^2+m^2})/m\,.
\end{equation} 

The localized state can be expanded in terms of the complete set of states $\psi_{\vec{k}}$ which are the solutions in the bulk.
Clearly any state which is localized at the impurity must be sensitive to the staggered field. This 
means that it can 
only contain bulk solutions which are substantial superpositions of $\vec{k}$ and $\vec{k}+\vec{\pi}$, i.e. 
$b_{\vec{k}}/a_{\vec{k}}\sim 1$. From the expression, Eq. \ref{bovera}, for $b_{\vec{k}}/a_{\vec{k}}$ we find
$\varepsilon^0_{\vec{k}}=\frac{1}{2}(a/b-b/a)m$  which implies that $\vec{k}$ is constrained to the
volume $|\varepsilon^0_{\vec{k}}|\alt m$.

\begin{figure}[h]
\includegraphics[scale=.4]{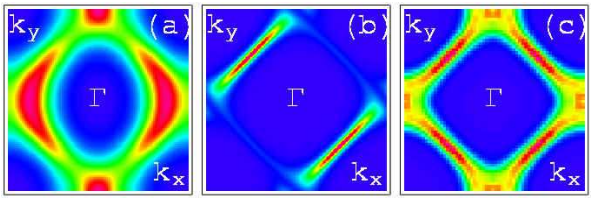}
\caption{\label{figfull_spectral} 
Spectral weight of various charge potentials. (a) is a stripe along the $y$ direction with $m=2,t'=-.1,t''=.1$, 
(b) is a diagonal stripe along $x+y$ direction with $m=0.5,t'=-.1,t''=0$ and (c) is a size $10\times 10$ bubble with 
$m=1,t'=-.1,t''=0$.}
\end{figure} 

In Fig. \ref{figfull_spectral} is shown the full spectral weight of the three types of potentials displayed in Fig. \ref{figvxy},
confirming our analytic result. 
The broader distribution in k-kspace for larger $m$ is consistent with a shorter localization length in real-space.  
Note also that for the 
diagonal stripe, Fig. \ref{figfull_spectral}b, the spectral weight is concentrated to those segments of the 
diamond which have momenta orthogonal to the stripe direction, a feature which will be important when analyzing diagonal 
stripes in the next section.

\section{Diagonal Stripes}

One of the most interesting recent stripe related findings is the diagonal stripes seen by quasi-elastic 
neutron scattering in very underdoped insulating phase of LSCO.\cite{Wakimoto99} 
As discussed in Sec. \ref{intro} there is a crossover region where both diagonal and 
bond-aligned stripes appear to coexist and which coincides with the rapid drop of the superconducting transition 
temperatures with decreasing doping,\cite{Fujita} pointing to a strong connection between stripes and superconductivity. 
On the other hand, we also noted that other properties, such as the nodal Fermi velocity and the hole mobility does not show any 
dramatic change across the bond-aligned to diagonal transition.

What we can contribute to this discussion within our model is a comparison between the expected distribution
of spectral weight between bond-aligned and diagonal stripes. It was already shown in Ref. \onlinecite{Salkola}
and \onlinecite{Granath} that a disordered bond-aligned stripe array can well reproduce the qualitative features of 
the near optimally doped samples. It is natural to do a similar analysis for the lightly doped samples 
$x=3-5\%$ studied by ARPES in Ref. \onlinecite{diagonal_LSCO_ARPES}. 
Experimentally it is found that the spectral weight in the anti-nodal, $(\pi,0)$,
regions of the BZ
which is most prominent for the under and optimally doped samples is gapped away from the Fermi energy, 
and instead the low-energy spectral
weight consists of disjoint arcs of ``Fermi surface'' centered near the nodal, $(\pi/2,\pi/2)$, regions. At a first glance
it is very tempting to identify these features with that shown in Fig. \ref{figfull_spectral}b for a diagonal
stripe. However, although it is probably not completely ruled out that this simple picture is correct 
a closer study of the diagonal stripe reveals a serious problem. The problem is that for $t'<0$, a 
less then half-filled, i.e. hole doped, diagonal stripe will have very little spectral weight in the nodal region, but only 
near the anti-nodal regions.  

\begin{figure}[h]
\includegraphics[scale=.4]{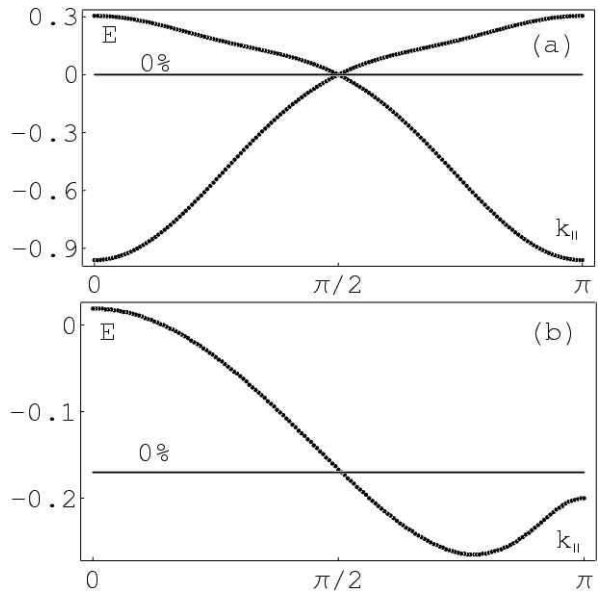}
\caption{\label{figsingle_dispersion} Dispersions of in-gap states for $m=t=1$, $t'=-.1$, and $t''=0$ for bond-aligned 
stripe (a) along $y$ with $k_{\|}=k_y$ and diagonal stripe (b) along $x+y$ with $k_{\|}=k_x+k_y$. Line marked ``0\%''
indicates no doped holes (half-filling).}
\end{figure}

In order to understand this statement we can consider the dispersion of the diagonal and bond-aligned stripe as a function of 
the conserved momentum along the stripe. We will restrict ourselves to the limit $m\rightarrow\infty$ and numerically 
confirm the qualitative correctness for $m=1$. In this limit it is trivial to solve for the spectrum 
on the stripe because the problem reduces to a one dimensional tight-binding chain. For the bond-aligned
stripe there is the nearest neighbor hopping $t$ while $t''$ acts as a next-nearest neighbor hopping, resulting in a 
dispersion
\begin{equation}
\varepsilon_{\text{col}}(k_{\|})\overset{m\rightarrow\infty}{=}-2t\cos(k_{\|})-2t''\cos(2k_{\|})\,.
\label{large_m_dispersion_col}
\end{equation}  
For the diagonal stripe the next-nearest neighbor hopping on the 2D lattice $t'$ acts as a nearest neighbor hopping on the
chain, giving a dispersion
\begin{equation}
\varepsilon_{\text{diag}}(k_{\|})\overset{m\rightarrow\infty}{=}-2t'\cos(k_{\|})\,.
\end{equation} 
In Fig. \ref{figsingle_dispersion} are shown numerically calculated dispersions for $m=1$ which agrees qualitatively 
with the large $m$ limit. Note that for the bond-aligned stripe there is a folding of the BZ due to the 
antiferromagnetic scattering along the stripe direction, which is absent for the diagonal stripe. The conclusion 
we want to draw from these dispersions is that for the diagonal stripe at any finite hole doping the momenta 
of filled states which may contribute to the low energy spectral weight are confined to $k_\|>\pi/2$. From the 
distribution of spectral weight for a diagonal stripe, Fig. \ref{figfull_spectral}b, we find that this implies
that there is very little spectral weight in the nodal regions.   
We can easily convince ourselves that this distribution of spectral weight of a diagonal stripe is a 
general consequence of the fact that a 
state with some momentum $k_\|$ will, as shown in Sec.\ref{sec:analytic}, have its spectral weight 
concentrated to the intersection of the line $k_x+k_y=k_\|$ with the $\cos(k_x)+\cos(k_y)=0$ diamond. This 
implies that for 
$|k_\||=\pi$ the weight will spread out over the whole intersection which is a line, while for 
$|k_\||<\pi$ the weight will be concentrated to the two points of intersection.

These results for the distribution of spectral weight of a hole doped diagonal stripe may be contrasted with that of a bond-aligned stripe.
Here we find from \ref{figsingle_dispersion}a complemented with Fig. \ref{figfull_spectral}a 
that for finite hole-doping the spectral weight may be spread out over the whole diamond. The details, of course,
depending on the doping, the parameters used and on the density of stripes. The latter being particularly important
in that for a short inter-stripe distance the stripe states will overlap and form bands of momenta transverse to 
stripes (see Fig. \ref{fig_inphase3}).  

One might, in an effective model such as this, attempt to fit the experimental data by taking $t'>0$
which would allow for spectral weight concentrated in the nodal region. 
However the ARPES data for the lightly doped samples indicate that most of the additional weight introduced with
doping is in fact in the anti-nodal region only that it is gapped away from the Fermi surface. If we take $t'>0$ 
we would vacate the anti-nodal states and we would not be able to reproduce this qualitative feature. 
Related to this there 
is a more quantitative problem for a diagonal stripe as contrasted with the ARPES data which should be more general
than our model, namely that the band width of a purely diagonal stripe is expected to be proportional to $t'$. 
Values of $t'$ in the literature is less than 0.1 eV, implying a band width $W_{\text{diagonal}}<.2$ eV, whereas the band width from the 
in-gap states seen in ARPES can be estimated at $W_{\text{in-gap}}\alt 1$ eV which looks more consistent with the band 
width $W_{\text{col}}\sim 2t$ of bond-aligned stripes. To summarize, we find that 
\textit{pure diagonal stripes are not consistent with the ARPES data of lightly doped LSCO}.

\subsection{Staircase Stripes}

\begin{figure}[h]
\includegraphics[scale=.4]{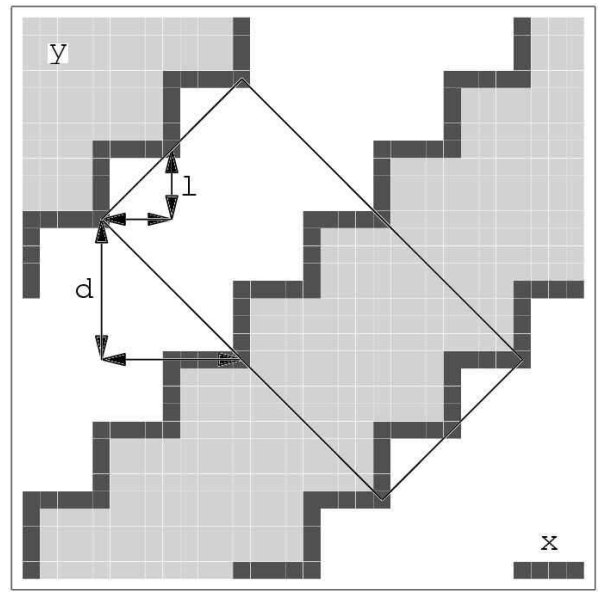}
\caption{\label{figstairs} Graphical representation of a staircase stripe defined by the 
step length $l$ and the stripe distance $d$. The rectangle indicates a primitive cell. 
The potential V(x,y) is given by V=0 for darker gray, V=1 for light gray, and V=-1 for white.}
\end{figure}

Given the difficulties with matching the model using a diagonal stripe configuration 
to the ARPES data it is instead tempting to 
look at bond-aligned stripes. Now, we know from neutron scattering that bond-aligned stripes are not seen 
in these very lightly doped materials, but only diagonal stripes. This led us to 
investigating the properties of stripes which are locally bond-aligned but globally diagonal. 
A natural and most simple candidate for such a construction is a ``staircase'' stripe. 
We can define a staircase stripe along the $x+y$ direction but letting it run alternately along the $\hat{x}$ and 
$\hat{y}$ directions with some step length $l$. For an ordered array of such staircase stripes we also introduce a 
stripe distance $d$ defined according to Fig. \ref{figstairs}. In the case of anti-phase stripes the potential has 
the symmetries 
\begin{eqnarray}
\label{stair_sym}
V(x+l,y+l)&=&V(x,y)\,,\nonumber\\ 
V(x+d,y-d)&=&-V(x,y)\,,
\end{eqnarray}
which also gives the primitive cell as indicated in the figure. The same symmetries hold true for the full potential 
of Eq. \ref{H} which is simply multiplied by a factor $(-1)^{x+y}$ to account for the staggered field. 
 
We will return to magnetic structure factors of such staircase stripes below, but it is easy to see that as long as
$l\leq d$ the main magnetic diffraction peaks of such a staircase stripe are equivalent to an array of purely diagonal 
stripes with the inter stripe distance $2d$ along the $x$ and $y$ direction.

\begin{figure}[h]
\includegraphics[scale=.4]{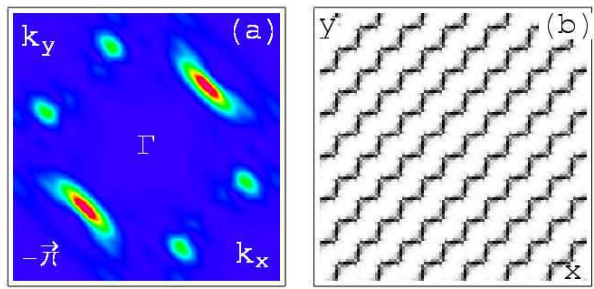}
\caption{\label{fig_stair2} 
Ordered array of staircase stripes with $l=8$ and $d=8$, using $t=m=1$, $t'=-.1$ and $t''=0$. 
Momentum space (a) and real space (b) spectral weight integrated over an energy window of $0.2$ around 
the Fermi energy at 4\% hole doping. The full system size is $256\times 256$ while in (b) is shown a 
$100\times 100$ section.}
\end{figure}

\begin{figure}[h]
\includegraphics[scale=.4]{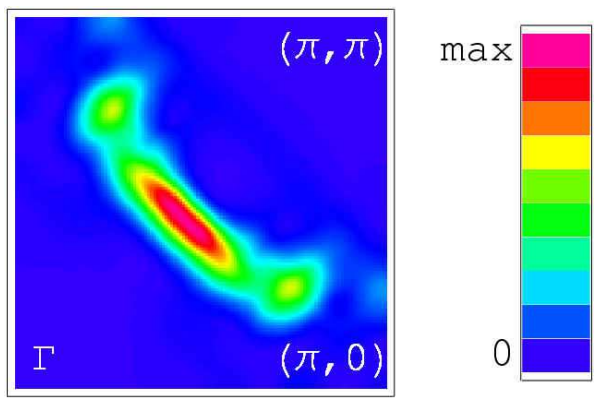}
\caption{\label{fig_stair3} 
Same as Fig. \ref{fig_stair2} but symmetrized with respect to the stripe orientation and in 
the first quadrant of the BZ. Shown is also the intensity map which has a linear scale.}
\end{figure}

We turn now to the distribution of spectral weight of staircase stripes. As an example we look at 
at a system with $l=8$ and $d=8$, were $d$ is chosen such that the magnetic structure factor has main peaks 
at $(\pi\pm\delta/\sqrt{2},\pi\mp\delta/\sqrt{2})$ with $\delta/\sqrt{2}=1/32$ which corresponds roughly to 
the $\delta\approx 1/25$ seen in Neutron scattering at 4\% doping. We diagonalize this system numerically to find
the single particle eigenstates $\psi_{\alpha}(\vec{k})$ with energies $E_\alpha$ in terms of which we 
calculate the single particle spectral function 
\begin{equation} 
A(\vec{k},\omega)=\sum_\alpha|\psi_\alpha(\vec{k})|^2\delta(E_\alpha-\omega)\,
\end{equation}
where $\delta(E_\alpha-\omega)$ is the Kronecker delta,
and the local density of states 
\begin{equation} 
R(\vec{r},\omega)=\sum_\alpha|\psi_\alpha(\vec{r})|^2\delta(E_\alpha-\omega)\,
\end{equation}
with $\psi_\alpha(\vec{r})=1/L_xL_y\sum_{\vec{k}}e^{i\vec{k}\cdot\vec{r}}\psi_\alpha(\vec{k})$ 
being the eigenfunctions in real space. 
In Fig. \ref{fig_stair2} is shown spectral weight distribution in 
$k$-space and real-space when integrated over an energy window $\Delta\omega=.2$ around the Fermi energy at 4\% doping,
i.e. calculating
\begin{equation}
I(\vec{k})=\int_{E_F-\Delta\omega/2}^{E_F+\Delta\omega/2}A(\vec{k},\omega)d\omega\,,
\end{equation}
and
\begin{equation}
R(\vec{r})=\int_{E_F-\Delta\omega/2}^{E_F+\Delta\omega/2}R(\vec{r},\omega)d\omega\,.
\end{equation} 
We find that the low-energy spectral weight is concentrated near the nodal region. It is highly anisotropic with most 
of the spectral weight parallel to the overall stripe direction along $x+y$ in contrast to the pure diagonal 
stripe shown in Fig. \ref{figfull_spectral}b. We have tried making the step length $l$ shorter, this gives results 
closer to the pure diagonal stripe with most of the spectral weight in the anti-nodal region. 
In Fig. \ref{fig_stair3} the results are symmetrized with respect to the 
stripe direction, so that $(\pi,\pi)=(\pi,-\pi)$ etc. Here the disjoint features merge into a single piece 
of ``Fermi arc'' in each quadrant of the BZ, similar to what is seen in ARPES.   
Note that in Fig. \ref{fig_stair2}b it is not the stripe potential $V(x,y)$ which is plotted, 
but the amplitude in real space of the low energy states. Not surprisingly these follow the potential 
quite closely, but there is some leakage of spectral weight into the antiferromagnetic regions which appears to smooth
the kinks and make the stripes more diagonal. An indication that the staircase stripes are just a caricature 
with the real stripes probably being smoother, but nevertheless locally closer to bond-aligned than to diagonal.   

\begin{figure}[h]
\includegraphics[scale=.4]{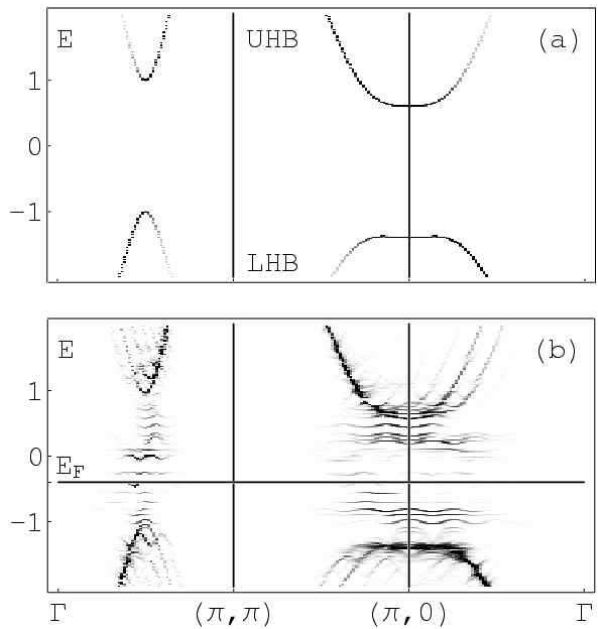}
\caption{\label{fig_stair4} Band structure of system with $t=m=1$, $t'=-.1$ and $t''=0$ for (a) no stripes and (b) with
staircase stripes as in Fig. \ref{fig_stair2}. UHB and LHB indicate upper and lower Hubbard bands respectively and 
$E_F$ is the Fermi energy at 4\% hole doping.  
The spectral weight is indicated by the intensity, but on a non-linear scale which exaggerates 
low-intensity features. (b) is symmetrized with respect to the stripe direction.}
\end{figure}

We can gain a better understanding of these results for the low-energy spectral weight by studying the spectral 
function, $A(\vec{k},\omega)$, over a broader energy window along the high-symmetry directions, as shown in 
Fig. \ref{fig_stair4}. 
The qualitative features of the spectral weight deriving from the staircase stripes can be directly linked to the 
properties of ordinary bond-aligned stripes. The high energy (away from $E_F$) spectral weight is concentrated around the
anti-nodal, $(\pi,0)$, region whereas the low energy spectral weight close to the Fermi energy is focused to the 
nodal region around $(\pi/2,\pi/2)$. This is what we would find from a lightly hole doped bond-aligned stripe with the 
dispersion in Fig. \ref{figsingle_dispersion}a, given the fact that the spectral weight is concentrated to the 
intersection of $k_\|=k_x$ or $k_y$ with the BZ diamond.

\subsubsection{Magnetic Structure factors of Staircase Stripes}

In the previous section we found a qualitative agreement of the spectral distribution of staircase stripes 
with that seen experimentally in lightly doped LSCO. Here we had to restrict ourselves to staircase stripes 
with step length $l$ less than the stripe distance $d$ as defined in Fig. \ref{figstairs} in order to have a 
magnetic structure factor which corresponds to diagonal stripes. A natural extension is to study also stairs with 
$l>d$. Certainly, in limit $l\gg d$ we expect very little influence from the kinks and the system will become 
equivalent to that of bond-aligned stripes, both for the spectral distribution and structure factors. 
We found that the distribution of spectral weight could be rationalized in terms of bond-aligned stripes already 
for the system with $l=d$ so that for $l>d$ we would not expect any qualitative difference between the 
spectral distribution of staircase stripes with that of bond-aligned stripes studied in the earlier 
work.\cite{Salkola,Granath} What is more interesting is to study the magnetic structure factor, which is 
sensitive to the global properties of the system. 

Physically, the relevant entity is the magnetic structure factor which is the amplitude $|S^z(\vec{q})|^2$ of 
of the Fourier transformed spin density which at zero temperature  reads
\begin{equation}
S^z(\vec{q})=1/L^2\sum_{\vec{r},\alpha,\sigma}e^{i\vec{q}\cdot\vec{r}}\sigma|\psi_{\alpha\sigma}(\vec{r})|^2
\Theta(E_F-E_\alpha)\,.
\end{equation}
However, for a large system, such as for disordered stripes, this is difficult to calculate because of the need 
to diagonalize the system.  
Much simpler to find is the amplitude squared of the Fourier transform of 
the stripy potential $(-1)^{x+y}V(x,y)$. We have checked for several ordered stripe arrays that close to 
half-filling the two calculations give very similar results. This is expected because the difference is roughly  
the amplitude on the stripes which is zero for the potential but slightly different from zero for the actual 
spin density because of the occupied stripe states.

\begin{figure}[h]
\includegraphics[scale=.4]{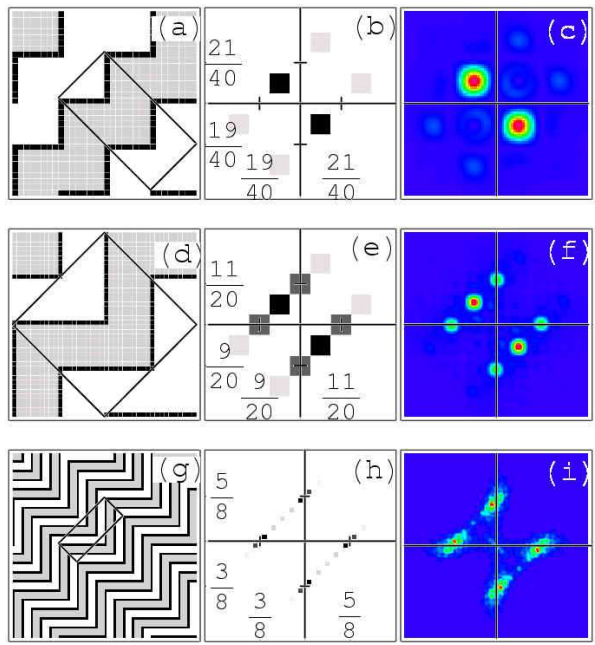}
\caption{\label{fig_stairsq} Magnetic structure factors of the three different regimes of staircase stripes discussed in text.
The left column shows the unit cells which in (a) is $80\times 80$ with step length $l=20$ and stripe distance 
$d=20$, in (d) is $40\times 40$ with $l=20$, $d=10$ and in 
(g) is $80\times 80$ with $l=20$, $d=4$ . 
Black are the stripes where the potential is zero, gray and white are the two $\pi$-shifted 
domains of the antiferromagnetic order, and the rectangles are primitive cells. The middle column, (b), (e) and (h),
shows the corresponding structure factors in reciprocal lattice units centered around $(1/2,1/2)=(\pi,\pi)$. The right 
column, (c), (f) and (i), give the structure factors of a sum over ten size $80\times 80$ disordered 
configurations of the 
corresponding ordered states to the left,
where the length of the legs and the distance between stripes is allowed to vary with a flat random distribution 
without allowing stripes to come closer 
than one site separation. The width of the peaks of the right column is simply related to the fraction of the BZ viewed, given 
that in all calculations the same correlation length is used.}
\end{figure}

Using the symmetry properties of a staircase stripe, Eq. \ref{stair_sym}, we find that $V(k_x,k_y)$ can 
only have non-zero components for 
\begin{eqnarray}
k_x+k_y&=&\frac{2\pi}{l}N,\quad N\in \text{Integer}\\
k_x-k_y&=&\frac{2\pi}{2d}N',\quad N'\in \text{odd Integer}\label{odd_V}\,.
\end{eqnarray}
In Fig. \ref{fig_stairsq} is shown the structure factors of three different stripe realizations, $l=d$, $l=2d$ and
$l=5d$. We find three very distinct diffraction patterns, were the first corresponds to diagonal stripes, the second
looks like diagonal stripes coexisting with both vertical and horizontal bond-aligned stripes, 
and the last  looks like vertical and horizontal stripes 
but with an orientation which deviates slightly from bond-aligned. We also show diffraction patterns
of samplings over disordered stripe configurations were we find that in general only the primary peaks survive with 
secondary, lower intensity peaks, getting washed out, although this does not happen for the second configuration 
with coexisting bond-aligned and diagonal peaks. We have also studied stripe arrays with larger unit cells where $l$
is not an integer factor of $d$. For disordered realizations it appears that these roughly fall into one of 
the three characteristic regimes, with only a quite narrow window of $l\approx 2d$ showing both diagonal and bond-aligned 
peaks of appreciable amplitude.

\begin{figure}[h]
\includegraphics[scale=.4]{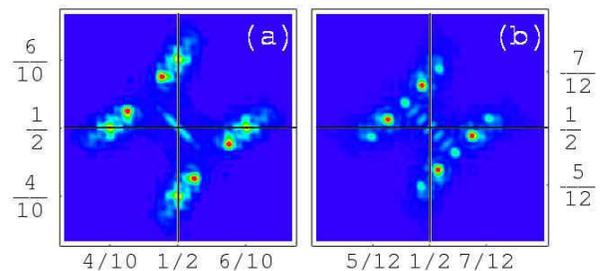}
\caption{\label{fig_stairs10} Magnetic structure factors of disordered realizations of staircase stripes with $l=20$ and 
(a) $d=5$ and (b) $d=6$ which could roughly correspond to 10\% and 8\% doping respectively. }
\end{figure}

These results for the diffraction pattern of staircase stripes appear qualitatively very similar to what is seen 
in neutron diffraction experiments on LSCO in the low-temperature orthorhombic (LTO) phase.\cite{Fujita} 
In the non-superconducting phase at very low doping ($x\alt 6\%$) is found 
peaks consistent with diagonal stripes, whereas at higher doping ($x\approx 10-13\%$) are found a pattern consistent
with stripes that are close to bond-aligned but shifted by a few degrees from the tetragonal axes. 
More recently was found that very close to the insulator to 
superconducting transition ($x\approx 6\%$) both patterns coexist. 
This may suggest that the \textit{stripes in the LTO phase of LSCO are always of the staircase type 
and that the transition between bond-aligned and diagonal stripes is a crossover from a regime where $l>2d$ to one where 
$l<2d$}. Particularly interesting would be to have more detailed experimental data from neutron scattering on the under 
doped superconducting regime $x\approx 10\%$. From the staircase stripe scenario we would expect a larger 
angle of deviation from the tetragonal axes than for the samples with higher doping, possibly together with secondary 
peaks in the diagonal direction or on the tetragonal axes as shown in Fig. \ref{fig_stairs10}.  

We should note that the correspondence between experiment and our results for staircase stripes is not perfect. In particular
there appears to be a discrepancy  with the relation between the ``incommensurabilty'' $\delta$, defined as the shift of a peak 
from $(\pi,\pi)$, of the diagonal and bond-aligned components.
For the staircase stripes we find $\delta_{\text{diag}}=\delta_{\text{col}}/\sqrt{2}$, whereas experimentally is 
seen $\delta_{\text{diag}}\approx\delta_{\text{col}}$. The former relation follows directly from the symmetry
of the stripy potential, Eq. \ref{odd_V}, and consequently is not sensitive to small changes of the potential
such as for instance smoothing of the kinks.

\section{In-phase Stripes}
We now turn to a study of the spectral weight distribution of in-phase stripes. As discussed in Sec. \ref{intro} 
there is indirect evidence for an inhomogeneous charge distribution in the electron doped PLCCO\cite{NMR_e_doped,Ando}. 
In fact, the possibility of in-phase stripes had already been suggested from theory for hole doped stripes at lower hole 
densities.\cite{Zachar}
 
\begin{figure}[h]
\includegraphics[scale=.4]{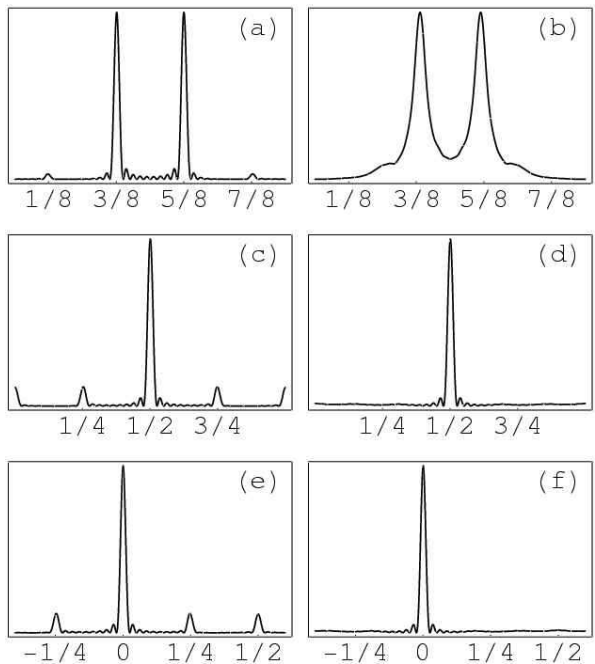}
\caption{\label{fig_stripes_sofq} Structure factors (in arbitrary units) of spin and charge density for anti-phase 
and in-phase stripes as a function of momentum transverse to the stripes in r.l.u.. 
The left column 
shows ordered arrays with stripe distance $d=4$ and the right disordered with  a flat distribution $d=2-6$. 
(a) and (b) shows the spin order for anti-phase stripes, (c) and (d) the spin order for in-phase stripes, and 
(e) and (f) the charge order which is independent of the type of stripe.
In all figures is used a finite correlation length of 40 sites and the disordered systems are averaged over 
1000 samples.}
\end{figure}

The experimental hallmark of anti-phase stripes is the incommensurate magnetism detected by neutron diffraction, 
where the weight is shifted away from the antiferromagnetic ordering vector $\vec{Q}=(\pi,\pi)$.  
A system with ordered in-phase stripes on the other hand will have a main diffraction peak at the 
antiferromagnetic ordering 
vector with satellites at positions shifted by integer multiples of $2\pi/d$, 
where $d$ is the distance between stripes. However, if the stripes are not static and ordered, but fluctuating or 
disordered, the satellite peaks will easily be washed out and the weight absorbed into the AF peak. In Fig. 
\ref{fig_stripes_sofq} is shown the amplitude squared of the Fourier transform of the stripe potential 
$(-1)^{x+y}V(x,y)$ along the transverse direction for ordered and disordered anti-phase 
and in-phase stripes and the for the charge density represented by $|V(x,y)|$.
For the charge order there is no difference between anti-phase and in-phase stripes. Nevertheless, charge 
order is more difficult to detect even for systems with static order. The reason for 
this is that neutrons do not couple directly to the charge order and that the  superlattice peaks related to
charge stripe order arise from modulations of the uniform charge density, $\vec{Q}=(0,0)$, 
which dominates the structure factor.  
As shown in Fig. \ref{fig_stripes_sofq} these peaks are also readily destroyed by disorder. (In fact, the charge 
structure factor as estimated here is equivalent to the spin structure factor of in-phase stripes 
but shifted by $\pi$.)  
Direct signatures of in-phase stripes from neutron or x-ray scattering will thus be much more difficult to find, 
most likely requiring a stripe ordered material.

\begin{figure}[h]
\includegraphics[scale=.4]{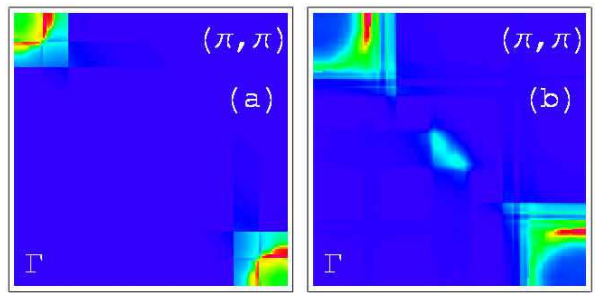}
\caption{\label{fig_inphase1} Spectral weight close to $E_F$ (window $.2$) 
of disordered in-phase stripes with $t=m=1$, $t'=-0.2$, 
and $t''=0.1$. Mean stripe distance in (a) is 12 with $E_F$ at 4\% doping and in (b) mean distance is 4 with 
$E_F$ at 10\% doping. The results are symmetrized with respect to the stripe direction. The system size is
$320\times 320$ .}
\end{figure} 

\begin{figure}[h]
\includegraphics[scale=.4]{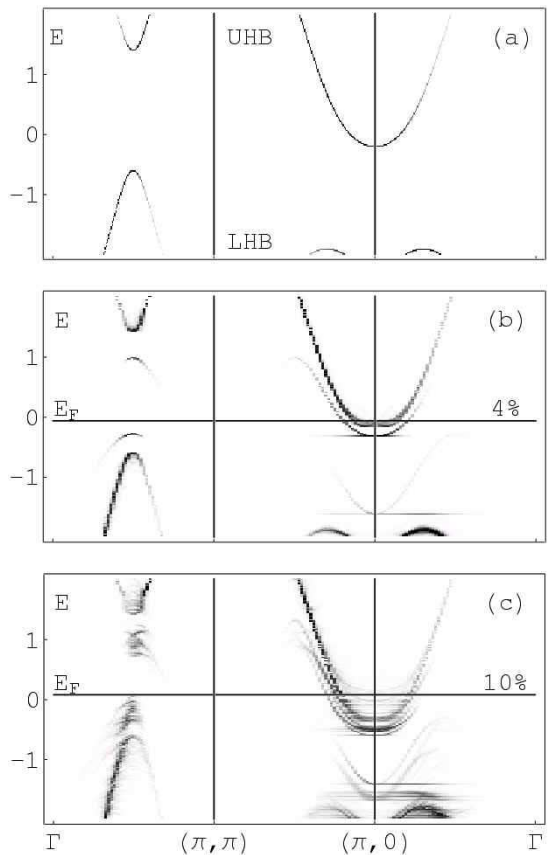}
\caption{\label{fig_inphase2} 
Band structure of system with $t=m=1$, $t'=-0.2$, and $t''=0.1$ and in-phase stripes as in Fig. 
\ref{fig_inphase2}. (a) is no stripes, (b) mean period 12 stripes and (c) mean period 4 stripes.}
\end{figure}

Nevertheless, given the indirect indications for the existence of in-phase stripes in 
electron doped cuprates it may be interesting 
to study the implications of such structures to the distribution of spectral weight. In Fig. \ref{fig_inphase1} is
shown the integrated spectral weight close to the Fermi energy for disordered in-phase stripes with 
inter-stripe distances 9-15 with mean 12 and 1-7 with mean 4 for a system 
with $t=m=1$, $t'=-.2$ and $t''=.1$. The parameters $t'$ and $t''$ are
chosen such as to roughly reproduce the band structure of the undoped system seen in ARPES on the electron doped 
NCCO\cite{e_doped_ARPES} with the smallest 
spectral gap at $(\pi/2,\pi/2)$ as shown in Fig. \ref{fig_inphase2}a. The density of stripes is assumed to be 
roughly as in the hole doped materials with stripe spacing $d$ given by the doping $n$ according to 
$n\approx 1/2d$, which for electron doping corresponds to $3/4$ filled stripes.

\begin{figure}[h]
\includegraphics[scale=.4]{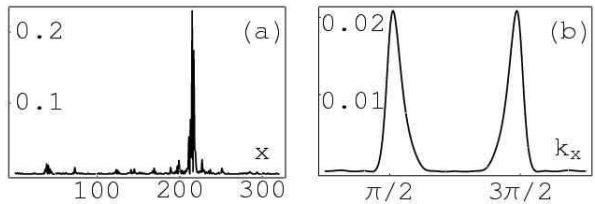}
\caption{\label{fig_local_state} Example of a single state $\psi$ of system in Fig. \ref{fig_inphase2}c, 
with energy $E\approx 0$ and momentum along stripe direction $k_y=\pi/2$, showing localization in the
transverse stripe direction. In (a) is shown the real space 
amplitude $|\psi(x,y)|^2$ as a function of transverse direction $x$ for arbitrary $y$ and in (b) 
the k-space amplitude $|\psi(k_x)|^2$. We can estimate a localization length $\xi\approx 2$ 
from the real space peak amplitude.}
\end{figure}

Clearly these results reproduce the experimental Fermi surface quite well, at light doping 
there are patches of spectral weight around $(\pi,0)$ while at higher doping the weight starts to look more like a full 
Fermi surface closed around $(\pi,\pi)$ but with weight missing at ``hot-spots'' where the putative Fermi surface cuts 
the BZ diagonal (diamond). Similar results have been reproduced by mean-field theory 
of the $t-t'-t''-U$ Hubbard model by allowing for a doping dependent $U$.\cite{Kusko} This is in fact the 
model we consider but without the stripes and with a magnitude of the staggered field $m$ which depends on $U$. 
Here, we keep the parameters 
fixed but vary the stripe spacing as a function of doping. There are however very distinct differences between
the implications of the two scenarios. In the stripe model there are mid gap states, not present for the model with 
a uniform staggered field. This is particularly visible for the lightly doped system, Fig. \ref{fig_inphase2}b, 
where along the $\Gamma=(0,0)$ to $(\pi,\pi)$ direction there is spectral weight between $E_F$ and the 
lower Hubbard band. Precisely such a feature is seen in the ARPES data for the $4\%$ doped sample.  
Secondly, the low-energy states of the stripe model are dynamically one-dimensional, i.e. they are localized 
transverse to a stripe but have a well defined momentum along the stripe, as shown in Fig. \ref{fig_local_state}. 
Thus, one of the main conclusions
is that \textit{the distribution of low-energy spectral weight, the 
``Fermi surface'', may be indistinguishable from a homogeneous two-dimensional system even though the 
states are dynamically one-dimensional}.

\begin{figure}[h]
\includegraphics[scale=.4]{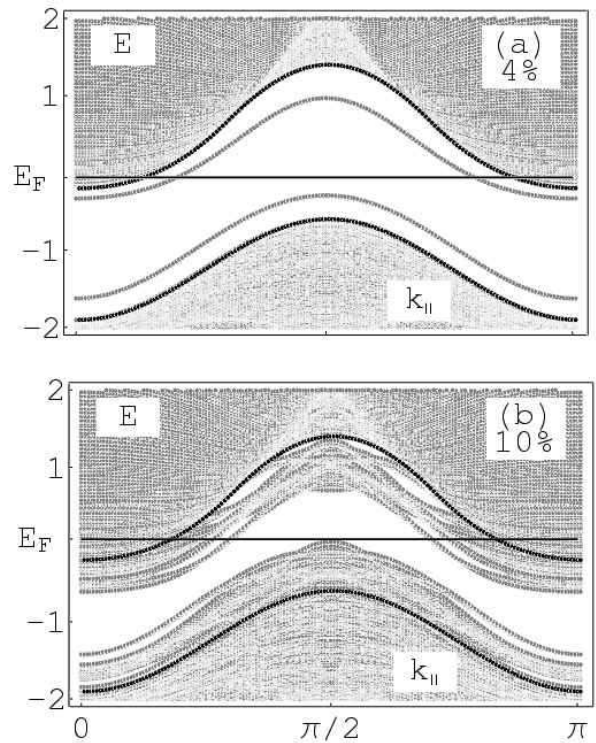}
\caption{\label{fig_inphase3} 
Spectrum for $|E|<2$ of in-phase stripe systems as in Fig. \ref{fig_inphase2} with mean stripe 
period 12 in (a) and 4 in (b) as a 
function of momentum $k\|=k_y$ parallel to the stripe direction $y$. The black curves indicate the bounds of the upper 
and lower Hubbard bands for the corresponding system without stripes.}
\end{figure}

Let us now look at how we can understand this evolution of spectral weight with doping and corresponding increasing
stripe density from the properties of a single in-phase stripe. Fig. \ref{fig_inphase3} shows the spectra of systems
with sparse, mean distance 12, and dense, mean distance 4, stripes. For the sparse stripes, Fig. \ref{fig_inphase3}a, 
the stripe states are 
clearly visible as the isolated in-gap bands. Because there is no broadening of the stripe states the stripes are 
clearly sufficiently far apart as to be effectively independent.   
We find that in-phase stripes at moderate $m$ have a band structure which is distinctly different
from that of anti-phase stripes as shown in Fig. \ref{figsingle_dispersion}a, with the former
staying close to the upper and lower Hubbard bands. (We have checked that in the limit 
$m\rightarrow\infty$, where in-phase and anti-phase stripes are equivalent, 
the dispersion Eq. \ref{large_m_dispersion_col} is correctly reproduced also for in-phase stripes.)  As the stripe
density is increased the stripe states will overlap and form bands transverse to the stripes, 
Fig. \ref{fig_inphase3}b, with states from the lower branch crossing the Fermi energy. Because of disorder, these 
states may nevertheless be strongly localized as shown in Fig. \ref{fig_local_state}. 

An important point about in-phase stripes which is demonstrated here is that the chemical potential may move as a 
function of doping and is not necessarily pinned 
within the gap. This is particularly evident in comparing the 0\% and 4\% samples, Fig. \ref{fig_inphase2}a and b, where in the former 
the chemical potential is in the gap ($-0.6<\mu<-0.2$) whereas in the latter it has moved up to cut the 
upper Hubbard band ($\mu\approx 0$). The motion of the chemical potential is a necessary consequence of the dispersion of in-phase stripes  
as shown in Fig. \ref{fig_inphase3}a, with the stripe states ``hugging'' the upper and lower Hubbard bands. This is in sharp contrast
to anti-phase stripes (Fig. \ref{fig_stair4} and \ref{figsingle_dispersion}), where the ``stripy'' spectral weight is ``mid-gap'' 
implying that the chemical potential may stay fixed with doping. \textit{Motion of the chemical potential with doping is thus not
necessarily an indication of the absence of stripes}.\cite{motion_doping,e_doped_ARPES}

\subsection{Bubbles}
Because of the lack of direct diffraction evidence for stripes one may be free to speculate on other forms of charge
order in these materials.
The microscopic motivation for stripe formation is the tendency of the antiferromagnet to expel extra charge 
which will disrupt the local antiferromagnetism. The formation of stripes is then a compromise between the minimizing the
magnetic energy by concentrating the holes, kinetic energy by allowing holes to delocalize along the stripes, and 
possibly the charging energy due to long range Coulomb repulsion by not allowing a macroscopic charge inhomogeneity.\cite{stripe_formation}

However, if the magnetism is relatively stronger than the kinetic energy contribution the system may prefer to 
keep the extra charge in zero-dimensional ``bubbles'' (see Fig. \ref{figvxy}c) in order to minimize the disruption of the local 
magnetic order.\cite{Steve_discuss}  
Roughly speaking, for a completely filled stripe with two electrons per site on a one-site wide stripe there are three
bad bonds where the spin exchange is destroyed per stripe site, while for a bubble there are only two bad bonds 
per site. (The sites on the perimeter of the bubble have three bad bonds).  
In addition, because of charging energy, the bubbles would have to be limited to a microscopic size.  
Clearly, the putative stripes or bubbles are not completely filled in the electron doped materials, 
due presumably to kinetic energy considerations, as this would imply an insulating system.

\begin{figure}[h]
\includegraphics[scale=.4]{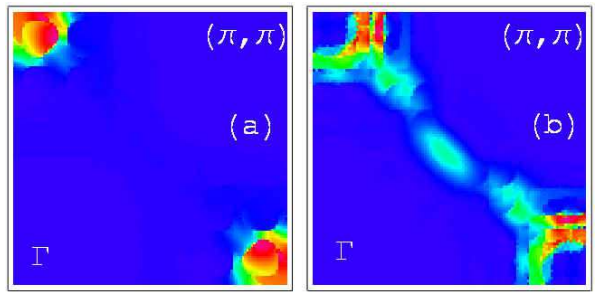}
\caption{\label{fig_blobFS} Spectral weight within energy window $.2$ of 
$E_F$ of ordered array of bubbles with $t=m=1$, $t'=-0.2$, 
and $t''=0.1$. In (a) the unit cell is $12\times 12$ with the bubble size 
$4\times 4$ and $E_F$ is at 4\% electron doping. 
In (b) the unit cell is $20\times 20$ with the bubble size $10\times 10$ and $E_F$ at 10\% doping.}
\end{figure} 

For simplicity we look at square bubbles and because this problem is fully two-dimensional with a large unit cell we
only consider ordered systems. For the previous study of in-phase stripes we chose stripe spacings and fillings which
were based on the corresponding values for anti-phase stripes in the hole doped systems. For the bubble phase we have 
even less to guide us on how to choose the size and period of the bubbles. 
However, assuming that the putative bubble formation is due to competition between the 
magnetic energy and the charging
energy we can get a simple estimate of the variation of bubble size with 
doping, ignoring in this completely the kinetic energy. It is easy to see that for a classical antiferromagnet 
we decrease the number of bad bonds which are not connecting 
antiferromagnetically aligned spins by two per doped electron by forming 
bubbles instead of a homogeneous (Wigner crystal) distribution of doped charge. This gives an energy gain 
$E_{\text{AF}}\sim -J L^2$, where $J$ is the AF exchange which is typically of the order of 
$0.1$ eV and $L$ is the 
linear dimension of the bubble. The charging energy due to moving $L^2$ 
electrons together from a closest distance $1/n$ to a closest distance $1$
is given by $E_{\text{Q}}\sim QL^3 (1-n)$ assuming a 
Coulomb interaction $V=Q/r$ with $r$ in units of the lattice constants. Here 
$Q=\frac{e^2}{4\pi\epsilon_0\epsilon a}\approx .3$ eV assuming a 
lattice constant of $a=5$\AA and a dielectric constant $\epsilon=10$.\cite{dielectric} 
Minimizing the total energy with respect to 
the bubble size $L$ gives $L\sim \frac{J}{Q(1-n)}$, so that for small doping,
$n$, we find the bubble size proportional to the doping. The prefactor $J/Q$ seems somewhat on the small side 
but does not rule out bubbles as a viable scenario. Clearly large $J$, or ``strong''
antiferromagnetism, will favor the bubble scenario. This being the reason why we suggest it for the electron 
doped materials which in general show antiferromagnetism over a much wider doping range than 
do hole doped materials. 
 
\begin{figure}[h]
\includegraphics[scale=.4]{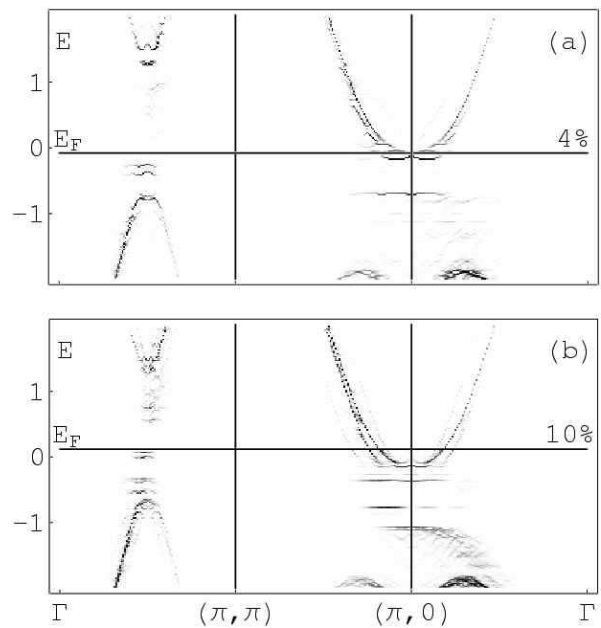}
\caption{\label{fig_blob_dispersion} 
Band structure of system in Fig. \ref{fig_blobFS}. The 
band structure without bubbles is the same as in Fig. \ref{fig_inphase2}a.}
\end{figure} 

For the calculations presented here we choose a bubble size $L=100n$
and let the distance between bubbles vary with doping such as to correspond to
roughly $3/4$ filled bubbles. In Fig. \ref{fig_blobFS} is shown the 
the integrated spectral weight near the Fermi energy for two realizations,  
at 4\% and 10\% doping, and in Fig \ref{fig_blob_dispersion} the 
corresponding band structure. 
The results are qualitatively similar to what was 
found for in-phase stripes. The evolution of the weight near $(\pi,0)$ does 
not depend crucially on the bubbles but can be understood from increasing 
filling of the upper Hubbard band. The nodal weight near $(\pi/2,\pi/2)$ 
on the other is clearly an effect of the additional states deriving from 
the bubbles.

\section{Conclusions}
We have investigated the distribution of electronic spectral weight in various charge ordered antiferromagnets.

We find that the spectral weight of states localized on a stripe or other charge structure is centered on and spread out over
the Brillouin zone diagonals ($\cos(k_x)+\cos(k_y)=0$).   
The distribution of spectral weight close to the Fermi energy may thus look fully two dimensional and practically indistinguishable
from a homogeneous system even though the low-energy states are dynamically one- or even zero-dimensional,
localized on stripes or bubbles. On the other hand the stripe states will in general lie within the energy gap of the undoped 
system making the appearance of ``in-gap'' states an expected consequence of an inhomogeneous charge distribution.

We find that pure diagonal stripes cannot reproduce the distribution of spectral weight found in ARPES on the very underdoped LSCO in 
the ``diagonal stripe'' phase. Instead we introduce ``staircase'' stripes which are locally vertical or horizontal but globally 
diagonal, in terms of which the qualitative features of the ARPES data is readily reproduced. Calculating the structure factors of
such staircase stripes we find that these evolve with doping and corresponding stripe density 
in a way which is very similar to the neutron scattering data in the 
LTO phase of LSCO over the whole doping range from very light to optimal. The results suggest that that the 
horizontal to diagonal stripe transition may be a crossover between a regime where the typical length of straight 
(horizontal or vertical) stripe segments  
is longer than the inter-stripe spacing to one where it is shorter and that the stripes are always locally bond-aligned in LSCO. 

We find that in-phase stripes can qualitatively reproduce the ARPES data end evolution with doping in 
the electron-doped cuprate NCCO. Particularly revealing is the spectral weight in the nodal region, near $(\pi/2,\pi/2)$,
where at low doping there is ``stripy'' spectral weight in the gap which at higher doping broadens as a result of increased stripe density 
and crosses the Fermi energy. For in-phase stripes, in contrast to anti-phase stripes, 
we find that the in-gap states lie close to the upper and lower Hubbard bands 
implying that the chemical potential is likely to move with doping.   
We also consider bubble structures and find that these produce similar 
results for the distribution of spectral weight to that of in-phase stripes by allowing the bubble size to grow with doping. 
We argue that these may be an alternative to stripes in the electron doped 
materials where there is a broad doping range with antiferromagnetic order. 

We want to thank S.A. Kivelson for valuable discussions. This work was supported by the Swedish Research Council.


\begin{thebibliography}{33}
\bibitem{bond_aligned} We will use the term ``bond-aligned stripes'' to indicate stripes which are aligned with the Cu-O bond direction
or the bonds of the square lattice in the model Hamiltonian. Such stripes are commonly refered to as vertical or horizontal but here we may 
or may not want to distinguish between the two.
\end{thebibliography}
\end{document}